\documentclass[conference,twocolumn]{IEEEtran}
\usepackage{graphicx}
\usepackage{amsmath}
\usepackage{amssymb}
\usepackage{cite}
\usepackage{hyperref}
\usepackage{tikz}
\usepackage{algorithm}
\usepackage{algorithmic}
\usepackage{float}
\usepackage{xcolor}
\usepackage[utf8]{inputenc}
\usepackage{hyperref}
\hypersetup{
  colorlinks=true,   % Set to true if you want colored text links
  linkcolor=blue,    % Color for internal links
  filecolor=magenta, % Color for file links
  urlcolor=blue,     % Color for external links
  citecolor=blue,    % Color for citations
  pdfborder={0 0 0}  % Removes the border around links
}
\begin{document}

\title{Blockchain Technology: Core Mechanisms, Evolution, and Future Implementation Challenges}

\author{
  \IEEEauthorblockN{Aditya Pratap Singh}
  \IEEEauthorblockA{
  National Institute of Technology, Tiruchirappalli\\
  Tiruchirappalli, India\\
  111122005@nitt.edu}
}

\maketitle

\begin{abstract}
Blockchain technology has emerged as one of the most transformative digital innovations of the 21st century. This paper presents a comprehensive review of blockchain's fundamental architecture, tracing its development from Bitcoin's initial implementation to current enterprise applications. We examine the core technical components including distributed consensus algorithms, cryptographic principles, and smart contract functionality that enable blockchain's unique properties. The historical progression from cryptocurrency-focused systems to robust platforms for decentralized applications is analyzed, highlighting pivotal developments in scalability, privacy, and interoperability. Additionally, we identify critical challenges facing widespread blockchain adoption, including technical limitations, regulatory hurdles, and integration complexities with existing systems. By providing this foundational understanding of blockchain technology, this paper contributes to ongoing research efforts addressing blockchain's potential to revolutionize data management across industries.
\end{abstract}


\section{Introduction}
Since the introduction of Bitcoin in 2008 \cite{nakamoto2008bitcoin}, blockchain technology has evolved from a niche cryptocurrency experiment into a revolutionary approach to digital record-keeping and transaction processing. At its core, blockchain represents a paradigm shift in how data is stored, validated, and shared across networks. By enabling secure, transparent, and immutable record-keeping without centralized authorities, blockchain technology has demonstrated potential to transform industries ranging from finance and supply chain management to healthcare and governance.

The fundamental innovation of blockchain lies in its unique combination of distributed consensus mechanisms, cryptographic techniques, and incentive structures that collectively create a system capable of establishing trust in trustless environments. Unlike traditional centralized databases managed by a single entity, blockchain distributes identical copies of a ledger across multiple participants in a network, ensuring that no single point of failure exists and that data remains accessible and verifiable by all authorized parties.

This paper aims to provide a comprehensive overview of blockchain technology, examining its core technical components, tracing its historical evolution, and identifying the key challenges that must be addressed for widespread adoption. We begin by exploring the fundamental mechanisms that enable blockchain's functionality, including distributed ledger systems, consensus algorithms, and cryptographic foundations. Next, we trace blockchain's development through distinct evolutionary phases, from cryptocurrency applications to programmable platforms and enterprise solutions. Finally, we discuss the significant technical, regulatory, and implementation challenges facing blockchain technology and potential approaches to addressing these limitations.

\section{Fundamental Blockchain Mechanisms}
\subsection{Distributed Ledger Architecture}
The foundation of blockchain technology is its distributed ledger architecture, which represents a fundamental departure from traditional centralized database systems. In a blockchain network, identical copies of the ledger are maintained across multiple nodes, creating a system that is resistant to single points of failure and censorship \cite{zheng2018blockchain}.

The distributed nature of blockchain provides several key advantages:

\begin{itemize}
    \item \textbf{Redundancy and Resilience}: Data is replicated across multiple nodes, ensuring continued operation even if individual nodes fail or are compromised.
    
    \item \textbf{Transparency}: All participants can view the same ledger, creating a shared source of truth accessible to all authorized parties.
    
    \item \textbf{Censorship Resistance}: The decentralized structure makes it extremely difficult for any single entity to alter historical records or prevent new transactions from being processed.
\end{itemize}

The blockchain ledger consists of blocks of data linked together in chronological order. Each block typically contains a batch of valid transactions, a timestamp, and a reference to the previous block (in the form of a cryptographic hash), creating a chain of blocks that extends back to the first block, known as the genesis block. This structure creates an immutable record of transactions, as altering any information would require modifying all subsequent blocks across the majority of nodes in the network—a task that becomes computationally infeasible as the chain grows longer.

\subsection{Consensus Mechanisms}
Consensus mechanisms are protocols that ensure all nodes in a blockchain network agree on the current state of the ledger without requiring trust between participants. These mechanisms represent one of the most significant innovations of blockchain technology, as they solve the Byzantine Generals Problem—a classic computer science challenge regarding the difficulty of reaching agreement in distributed systems when some participants may be unreliable or malicious \cite{lamport1982byzantine}.

Several consensus algorithms have been developed for blockchain systems, each with unique characteristics and trade-offs:

\subsubsection{Proof of Work (PoW)}
First implemented in Bitcoin, Proof of Work requires participants (miners) to solve computationally intensive mathematical puzzles to validate transactions and create new blocks \cite{nakamoto2008bitcoin}. The first miner to solve the puzzle broadcasts the solution to the network, allowing other nodes to verify its correctness. Once verified, the new block is added to the chain, and the miner receives a reward.

PoW provides strong security guarantees, as altering historical data would require controlling more than 50\% of the network's computational power (known as a 51\% attack). However, this security comes at the cost of high energy consumption and limited scalability, as the system can only process a finite number of transactions per block, and blocks are created at fixed intervals.

\subsubsection{Proof of Stake (PoS)}
Proof of Stake addresses the energy consumption concerns of PoW by selecting validators based on the amount of cryptocurrency they hold and are willing to "stake" as collateral, rather than computational power \cite{king2012ppcoin}. Validators are chosen to create new blocks according to various selection methods, including random selection weighted by stake size.

If validators attempt to validate fraudulent transactions, they risk losing their staked assets, creating an economic incentive for honest behavior. PoS systems consume significantly less energy than PoW while maintaining security through economic incentives rather than computational puzzles.

\subsubsection{Delegated Proof of Stake (DPoS)}
Delegated Proof of Stake extends the PoS concept by allowing stakeholders to vote for a small number of delegates who are responsible for validating transactions and creating blocks \cite{larimer2014delegated}. This approach improves scalability by reducing the number of nodes that actively participate in consensus, enabling faster transaction processing.

DPoS sacrifices some degree of decentralization for improved performance, as power is concentrated among a relatively small group of delegates. However, the voting mechanism ensures that delegates who act maliciously or inefficiently can be replaced, maintaining accountability within the system.

\subsubsection{Practical Byzantine Fault Tolerance (PBFT)}
PBFT is a consensus algorithm designed for permissioned blockchain networks, where participants are known and authorized \cite{castro1999practical}. Unlike PoW and PoS, which are probabilistic consensus mechanisms, PBFT provides deterministic finality, meaning that once a transaction is confirmed, it cannot be reversed.

PBFT operates through a multi-round voting process among validator nodes, requiring at least two-thirds of nodes to agree on the state of the ledger. This approach offers high transaction throughput and energy efficiency but requires a relatively small number of trusted validator nodes, making it suitable for enterprise applications rather than public, permissionless networks.

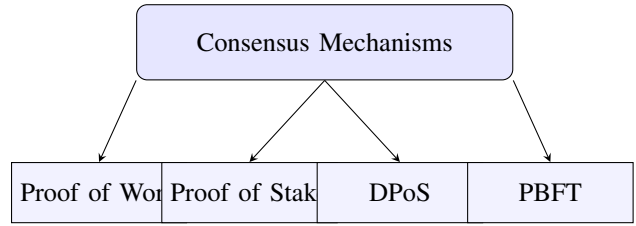
\begin{figure}[h]
\centering
\begin{tikzpicture}
% Define consensus types
\node[draw, rectangle, rounded corners, minimum width=5cm, minimum height=1cm, fill=blue!10] (consensus) at (0,0) {Consensus Mechanisms};

% Mechanism types with explicit coordinates
\node[draw, rectangle, minimum width=2.2cm, minimum height=0.8cm, fill=blue!5] (pow) at (-3,-2) {Proof of Work};
\node[draw, rectangle, minimum width=2.2cm, minimum height=0.8cm, fill=blue!5] (pos) at (-1,-2) {Proof of Stake};
\node[draw, rectangle, minimum width=2.2cm, minimum height=0.8cm, fill=blue!5] (dpos) at (1,-2) {DPoS};
\node[draw, rectangle, minimum width=2.2cm, minimum height=0.8cm, fill=blue!5] (pbft) at (3,-2) {PBFT};

% Connect with arrows
\draw[-stealth] (consensus.south west) -- (pow.north);
\draw[-stealth] (consensus.south) to[xshift=-1cm] (pos.north);
\draw[-stealth] (consensus.south) to[xshift=1cm] (dpos.north);
\draw[-stealth] (consensus.south east) -- (pbft.north);
\end{tikzpicture}
\caption{Major Blockchain Consensus Mechanisms}
\label{fig:consensus}
\end{figure}

\subsection{Cryptographic Foundations}
Cryptography forms the backbone of blockchain technology, providing the mechanisms for secure data storage, transaction verification, and user authentication. Several cryptographic techniques are fundamental to blockchain operations:

\subsubsection{Hash Functions}
Cryptographic hash functions transform input data of any size into a fixed-size output (hash) that uniquely represents the original data \cite{antonopoulos2014mastering}. In blockchain systems, hash functions serve multiple purposes:

\begin{itemize}
    \item Creating unique identifiers for blocks
    \item Linking blocks together in the chain (each block contains the hash of the previous block)
    \item Generating addresses for user accounts
    \item Verifying data integrity
    \item Providing computational puzzles for Proof of Work consensus
\end{itemize}

Common hash functions used in blockchain include SHA-256 (used in Bitcoin) and Keccak-256 (used in Ethereum). These functions are designed to be one-way (it is computationally infeasible to derive the input from the output) and collision-resistant (it is extremely unlikely for two different inputs to produce the same output).

\subsubsection{Public-Key Cryptography}
Public-key cryptography, also known as asymmetric cryptography, uses pairs of keys: a public key that can be shared openly and a corresponding private key that must be kept secret. This system enables two critical blockchain functions:

\begin{itemize}
    \item \textbf{Digital Signatures}: Users sign transactions with their private keys, and these signatures can be verified by anyone using the corresponding public keys. This mechanism ensures that only the rightful owner of assets can authorize their transfer.
    
    \item \textbf{Address Generation}: Public keys (or derivatives thereof) are used to generate blockchain addresses, allowing users to receive assets without revealing their full public keys until they spend from those addresses.
\end{itemize}

The security of public-key cryptography relies on mathematical problems that are computationally difficult to solve, such as the discrete logarithm problem (used in ECDSA, the signature algorithm used in many blockchains) or the factorization of large prime numbers (used in RSA).

\subsubsection{Merkle Trees}
Merkle trees are data structures that efficiently summarize and verify the integrity of large datasets \cite{merkle1987digital}. In blockchain systems, they are used to organize transactions within blocks, enabling quick verification of whether a particular transaction is included in a block without downloading the entire block.

Each leaf node of a Merkle tree represents the hash of a transaction, and each non-leaf node represents the hash of its child nodes. The top-level hash (the Merkle root) is included in the block header, allowing verification of any transaction by generating a Merkle proof—a minimal path of hashes required to compute the Merkle root from a particular transaction.

This structure provides significant efficiency benefits for lightweight clients (such as mobile wallets) that need to verify specific transactions without storing the entire blockchain.

\section{Historical Evolution of Blockchain Technology}

\subsection{Generation 1.0: Bitcoin and Cryptocurrency (2008-2013)}
The first practical implementation of blockchain technology came with the introduction of Bitcoin in 2009, following the publication of Satoshi Nakamoto's whitepaper titled "Bitcoin: A Peer-to-Peer Electronic Cash System" \cite{nakamoto2008bitcoin}. This groundbreaking work outlined a solution to the double-spending problem—preventing the same digital currency from being spent multiple times—without requiring a trusted third party.

Bitcoin demonstrated the potential of blockchain as a decentralized payment system by combining several existing technologies in a novel way:

\begin{itemize}
    \item A distributed ledger maintained by a peer-to-peer network
    \item Proof of Work consensus to validate transactions and secure the network
    \item Cryptographic techniques for transaction authentication and verification
    \item Economic incentives (block rewards and transaction fees) to encourage participation
\end{itemize}

Following Bitcoin's introduction, numerous alternative cryptocurrencies (altcoins) emerged, each introducing modifications to Bitcoin's original design. Notable examples include:

\begin{itemize}
    \item \textbf{Litecoin (2011)}: Featuring faster block generation times and a different hashing algorithm (Scrypt instead of SHA-256)
    
    \item \textbf{Ripple (2012)}: Designed for faster settlement in financial institutions, using a consensus protocol different from Bitcoin's Proof of Work
    
    \item \textbf{Monero (2014)}: Focusing on privacy features through ring signatures and stealth addresses
\end{itemize}

This first generation of blockchain technology focused primarily on peer-to-peer payments and store of value applications, with limited programmability beyond basic transaction types.

\subsection{Generation 2.0: Smart Contracts and Decentralized Applications (2014-2017)}
The second generation of blockchain technology emerged with the launch of Ethereum in 2015, following Vitalik Buterin's whitepaper published in late 2013 \cite{buterin2014ethereum}. While Bitcoin focused on a specific application (peer-to-peer electronic cash), Ethereum introduced a general-purpose blockchain platform capable of executing arbitrary code in the form of smart contracts.

Smart contracts are self-executing agreements with the terms directly written into code. They run on the blockchain, automatically enforcing their conditions when triggered, without requiring trusted intermediaries. This innovation expanded blockchain's potential beyond simple value transfer to more complex applications across various domains.

Key features of Generation 2.0 blockchains include:

\begin{itemize}
    \item \textbf{Turing-complete Programming}: The ability to express any computable function, enabling complex application logic
    
    \item \textbf{Decentralized Applications (dApps)}: Software applications running on a blockchain network rather than a single centralized server
    
    \item \textbf{Tokens and Tokenization}: The ability to create and manage digital assets beyond the native cryptocurrency
    
    \item \textbf{Decentralized Autonomous Organizations (DAOs)}: Organizations governed by smart contracts rather than traditional hierarchical structures
\end{itemize}

The introduction of smart contract platforms led to an explosion of innovation, including the rise of Initial Coin Offerings (ICOs) as a fundraising mechanism, decentralized finance (DeFi) applications, and non-fungible tokens (NFTs). However, this generation also exposed limitations in blockchain scalability, as evidenced by network congestion and high transaction fees during periods of high demand.

\subsection{Generation 3.0: Scaling, Interoperability, and Enterprise Adoption (2017-Present)}
The third generation of blockchain technology has focused on addressing the limitations of earlier systems, particularly regarding scalability, interoperability, and governance. Key developments in this generation include:

\subsubsection{Scalability Solutions}
To overcome the throughput limitations of earlier blockchains, several approaches have been developed:

\begin{itemize}
    \item \textbf{Layer 2 Solutions}: Protocols built on top of existing blockchains to handle transactions off the main chain, such as Lightning Network for Bitcoin and Optimistic Rollups for Ethereum
    
    \item \textbf{Sharding}: Partitioning the blockchain into smaller pieces (shards) that can process transactions in parallel
    
    \item \textbf{Alternative Consensus Mechanisms}: Moving from energy-intensive Proof of Work to more efficient mechanisms like Proof of Stake
    
    \item \textbf{New Blockchain Architectures}: Platforms designed for high throughput from the ground up, such as Solana and Avalanche
\end{itemize}

\subsubsection{Interoperability Frameworks}
As the blockchain ecosystem has grown more diverse, the need for different networks to communicate and share data has become increasingly important. Projects addressing this challenge include:

\begin{itemize}
    \item \textbf{Cross-Chain Protocols}: Systems like Polkadot and Cosmos that enable communication between independent blockchains
    
    \item \textbf{Atomic Swaps}: Cryptographic techniques for trustless exchange of assets across different blockchains
    
    \item \textbf{Wrapped Tokens}: Representations of assets from one blockchain on another blockchain
\end{itemize}

\subsubsection{Enterprise and Permissioned Blockchains}
The third generation has also seen increased adoption of blockchain technology by enterprises and institutions, often through permissioned networks that restrict participation to authorized entities:

\begin{itemize}
    \item \textbf{Hyperledger}: An umbrella project of open-source blockchains and related tools designed for enterprise use
    
    \item \textbf{Enterprise Ethereum}: Adaptations of Ethereum technology for business applications with privacy and scalability enhancements
    
    \item \textbf{Consortium Blockchains}: Networks operated by groups of organizations within specific industries, such as R3's Corda in financial services
\end{itemize}

These developments have expanded blockchain applications beyond cryptocurrency to areas such as supply chain management, digital identity, healthcare records, and government services.

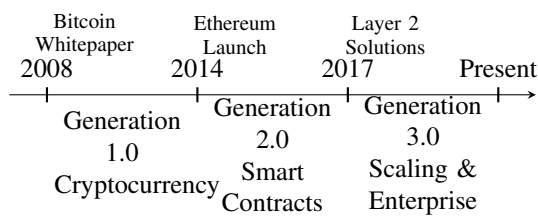
\begin{figure}[h]
\centering
\begin{tikzpicture}
% Define timeline
\draw[-stealth] (-0.5,0) -- (6.5,0);

% Timeline periods
\draw[thick] (0,-0.1) -- (0,0.1) node[above] {2008};
\draw[thick] (2,-0.1) -- (2,0.1) node[above] {2014};
\draw[thick] (4,-0.1) -- (4,0.1) node[above] {2017};
\draw[thick] (6,-0.1) -- (6,0.1) node[above] {Present};

% Era designations
\node[align=center, text width=1.8cm] at (1,-0.8) {Generation 1.0\\Cryptocurrency};
\node[align=center, text width=1.8cm] at (3,-0.8) {Generation 2.0\\Smart Contracts};
\node[align=center, text width=1.8cm] at (5,-0.8) {Generation 3.0\\Scaling \& Enterprise};

% Key events
\node[align=center, font=\footnotesize, text width=1.5cm] at (0.5,0.8) {Bitcoin Whitepaper};
\node[align=center, font=\footnotesize, text width=1.5cm] at (2.5,0.8) {Ethereum Launch};
\node[align=center, font=\footnotesize, text width=1.5cm] at (4.5,0.8) {Layer 2 Solutions};

\end{tikzpicture}
\caption{Evolution of Blockchain Technology}
\label{fig:evolution}
\end{figure}

\section{Current Applications and Use Cases}
\subsection{Financial Services}
The financial sector has been at the forefront of blockchain adoption, with applications ranging from cryptocurrency trading to complex decentralized finance (DeFi) protocols. Key use cases include:

\begin{itemize}
    \item \textbf{Cross-Border Payments}: Blockchain networks like Ripple and Stellar facilitate faster and cheaper international transfers compared to traditional banking systems.
    
    \item \textbf{Decentralized Finance (DeFi)}: Protocols offering lending, borrowing, trading, and insurance services without intermediaries, primarily built on smart contract platforms like Ethereum.
    
    \item \textbf{Asset Tokenization}: Converting real-world assets such as real estate, art, and commodities into digital tokens on a blockchain, enabling fractional ownership and improved liquidity.
    
    \item \textbf{Central Bank Digital Currencies (CBDCs)}: Digital versions of national currencies issued by central banks, potentially using blockchain or distributed ledger technology.
\end{itemize}

\subsection{Supply Chain Management}
Blockchain technology enables transparent and immutable tracking of products throughout supply chains, addressing challenges related to provenance, counterfeit prevention, and regulatory compliance:

\begin{itemize}
    \item \textbf{Product Provenance}: Tracking the origin and journey of products, particularly valuable for items like diamonds, luxury goods, and pharmaceuticals.
    
    \item \textbf{Food Safety}: Tracing agricultural products from farm to table, enabling quick identification of contamination sources during recalls.
    
    \item \textbf{Logistics Optimization}: Improving coordination between supply chain participants through shared, real-time visibility of shipment status and documentation.
\end{itemize}

IBM Food Trust, TradeLens, and VeChain are examples of blockchain platforms designed specifically for supply chain applications.

\subsection{Healthcare}
In healthcare, blockchain can address challenges related to data security, interoperability, and patient consent management:

\begin{itemize}
    \item \textbf{Medical Records}: Providing secure, patient-controlled access to health records across different providers.
    
    \item \textbf{Pharmaceutical Supply Chain}: Tracking medications from manufacturer to patient to combat counterfeit drugs.
    
    \item \textbf{Clinical Trials}: Improving data integrity and participant consent management in research studies.
\end{itemize}

Projects like MedRec and Patientory are exploring blockchain applications in healthcare data management.

\subsection{Identity and Governance}
Blockchain-based identity systems offer potential solutions to privacy concerns, identity theft, and exclusion from financial services:

\begin{itemize}
    \item \textbf{Self-Sovereign Identity}: Systems allowing individuals to control their personal data and selectively share verified credentials without revealing unnecessary information.
    
    \item \textbf{Voting Systems}: Secure, transparent election platforms that maintain voter privacy while preventing fraud.
    
    \item \textbf{Public Records}: Immutable registers for land titles, business licenses, and other government-issued certifications.
\end{itemize}

\section{Challenges and Future Directions}

\subsection{Technical Challenges}
\subsubsection{The Scalability Trilemma}
Blockchain systems face fundamental trade-offs between decentralization, security, and scalability—a challenge often referred to as the "blockchain trilemma" \cite{vitalik2021trilemma}. Most current systems optimize for two of these properties at the expense of the third:

\begin{itemize}
    \item Public blockchains like Bitcoin prioritize security and decentralization but have limited throughput.
    
    \item Permissioned networks achieve higher scalability by sacrificing some degree of decentralization.
    
    \item Some high-throughput public chains achieve scalability through more centralized validation mechanisms.
\end{itemize}

Ongoing research into layer 2 solutions, sharding, and novel consensus mechanisms aims to address this trilemma, but finding the optimal balance remains a significant challenge.

\subsubsection{Energy Consumption}
The energy consumption of Proof of Work blockchains has raised environmental concerns, with Bitcoin mining alone consuming more electricity than some countries \cite{de2018bitcoin}. While the transition to Proof of Stake significantly reduces energy requirements, this approach introduces its own challenges related to stake distribution and potential centralization.

\subsubsection{Privacy and Confidentiality}
Most public blockchains provide pseudonymity rather than true privacy, as all transaction data is visible on the public ledger. This transparency, while beneficial for certain applications, poses challenges for use cases requiring confidentiality:

\begin{itemize}
    \item \textbf{Zero-Knowledge Proofs}: Cryptographic techniques that allow verification of information without revealing the information itself
    
    \item \textbf{Private Transactions}: Protocols like Zcash and Monero that hide transaction details while maintaining verifiability
    
    \item \textbf{Confidential Smart Contracts}: Systems that execute contract logic without revealing the underlying data to network participants
\end{itemize}

\subsection{Regulatory and Legal Challenges}
The decentralized and borderless nature of blockchain technology presents unique regulatory challenges. Different jurisdictions have adopted varying approaches to blockchain regulation, creating a complex landscape for global operations:

\begin{itemize}
    \item \textbf{Regulatory Uncertainty}: Unclear or rapidly changing regulations discourage investment and adoption.
    
    \item \textbf{Compliance Requirements}: Anti-money laundering (AML) and know-your-customer (KYC) regulations may conflict with blockchain's pseudonymous design.
    
    \item \textbf{Legal Status of Smart Contracts}: Questions about the enforceability and legal recognition of automated agreements.
    
    \item \textbf{Data Protection Laws}: Tensions between immutable blockchain records and regulations like GDPR's "right to be forgotten."
\end{itemize}

\subsection{Implementation and Adoption Challenges}
Beyond technical and regulatory concerns, practical challenges to blockchain adoption include:

\begin{itemize}
    \item \textbf{Integration with Legacy Systems}: Connecting blockchain networks with existing enterprise software and databases.
    
    \item \textbf{Standards and Interoperability}: Lack of common standards hampering communication between different blockchain platforms.
    
    \item \textbf{User Experience}: Complex interfaces and key management requirements creating barriers to non-technical users.
    
    \item \textbf{Governance Structures}: Establishing effective decision-making processes for protocol upgrades and dispute resolution.
\end{itemize}

\subsection{Future Research Directions}
Key areas for future blockchain research include:

\begin{itemize}
    \item \textbf{Quantum-Resistant Cryptography}: Developing cryptographic techniques secure against quantum computing attacks.
    
    \item \textbf{Formal Verification}: Creating methods to mathematically prove the correctness of smart contracts before deployment.
    
    \item \textbf{Scalable Consensus}: Continuing research into consensus mechanisms that maintain security while improving throughput.
    
    \item \textbf{Cross-Chain Communication}: Developing secure and efficient protocols for interoperability between different blockchain networks.
    
    \item \textbf{Sustainable Blockchain Design}: Creating environmentally friendly systems without compromising security or decentralization.
\end{itemize}

\section{Conclusion}
Blockchain technology has evolved significantly since its inception with Bitcoin, expanding from a cryptocurrency innovation to a versatile platform for decentralized applications across various industries. The core principles of distributed ledgers, consensus mechanisms, and cryptographic verification continue to drive blockchain's potential to transform how we manage and share digital information.

Despite the considerable progress, blockchain technology still faces substantial challenges in terms of scalability, energy efficiency, privacy, regulation, and practical implementation. Addressing these challenges requires collaborative efforts from researchers, developers, industry stakeholders, and regulators.

As research continues and new solutions emerge, blockchain technology is likely to find increasing adoption in applications where its unique properties—transparency, immutability, and decentralization—provide significant advantages over traditional centralized systems. By understanding both the potential and limitations of blockchain technology, researchers and practitioners can contribute to its responsible development and implementation across various domains.

The future of blockchain will likely involve a diverse ecosystem of interconnected networks with different design priorities, ranging from highly decentralized public blockchains to performance-optimized enterprise solutions. This diversity, coupled with ongoing technical innovation, suggests that blockchain technology will continue to evolve and expand its impact on digital infrastructure in the coming years.


\begin{thebibliography}{00}
\bibitem{nakamoto2008bitcoin} S. Nakamoto, "Bitcoin: A Peer-to-Peer Electronic Cash System," 2008. [Online]. Available: https://bitcoin.org/bitcoin.pdf

\bibitem{zheng2018blockchain} Z. Zheng, S. Xie, H. Dai, X. Chen, and H. Wang, "Blockchain challenges and opportunities: A survey," Int. J. Web Grid Serv., vol. 14, no. 4, pp. 352–375, 2018.

\bibitem{lamport1982byzantine} L. Lamport, R. Shostak, and M. Pease, "The Byzantine Generals Problem," ACM Trans. Program. Lang. Syst., vol. 4, no. 3, pp. 382–401, 1982.

\bibitem{king2012ppcoin} S. King and S. Nadal, "PPCoin: Peer-to-Peer Crypto-Currency with Proof-of-Stake," 2012. [Online]. Available: https://peercoin.net/assets/paper/peercoin-paper.pdf

\bibitem{larimer2014delegated} D. Larimer, "Delegated Proof-of-Stake (DPOS)," Bitshares whitepaper, 2014.

\bibitem{castro1999practical} M. Castro and B. Liskov, "Practical Byzantine Fault Tolerance," in Proceedings of the Third Symposium on Operating Systems Design and Implementation, 1999, pp. 173–186.

\bibitem{antonopoulos2014mastering} A. M. Antonopoulos, "Mastering Bitcoin: Unlocking Digital Cryptocurrencies," O'Reilly Media, Inc., 2014.

\bibitem{merkle1987digital} R. C. Merkle, "A Digital Signature Based on a Conventional Encryption Function," in Advances in Cryptology — CRYPTO '87, 1987, pp. 369–378.

\bibitem{buterin2014ethereum} V. Buterin, "Ethereum: A Next-Generation Smart Contract and Decentralized Application Platform," 2014. [Online]. Available: https://ethereum.org/en/whitepaper/

\bibitem{vitalik2021trilemma} V. Buterin, "Why sharding is great: demystifying the technical properties," Ethereum Blog, 2021.

\bibitem{de2018bitcoin} A. de Vries, "Bitcoin's Growing Energy Problem," Joule, vol. 2, no. 5, pp. 801–805, 2018.

\bibitem{wang2019survey} W. Wang et al., "A Survey on Consensus Mechanisms and Mining Strategy Management in Blockchain Networks," IEEE Access, vol. 7, pp. 22328–22370, 2019.

\bibitem{chen2018survey} W. Chen et al., "A Survey on Blockchain Applications in Different Domains," in Proceedings of the 2018 International Conference on Blockchain Technology and Application, 2018, pp. 17–21.

\bibitem{tapscott2016blockchain} D. Tapscott and A. Tapscott, "Blockchain Revolution: How the Technology Behind Bitcoin Is Changing Money, Business, and the World," Portfolio, 2016.

\end{thebibliography}
\end{document}